\documentclass[%
 reprint,
 amsmath,amssymb,
 aps,
 prl,
 longbibliography,
 lengthcheck,%
]{revtex4-1}

\usepackage{graphicx}
\usepackage{dcolumn}
\usepackage{bm}
\usepackage{hyperref}

\begin{document}

\preprint{APS/123-QED}

\title{ Near-linear Dynamics for Shallow Water Waves}

\author{ M.B.~Erdo\u gan, N.~Tzirakis, V.~Zharnitsky  }
\affiliation
 {Department of Mathematics,
         University of Illinois at Urbana-Champaign,
                                   Urbana, IL 61801  }

\date{\today}

\begin{abstract}
It is shown that spatially periodic one-dimensional surface waves in shallow water behave
almost linearly,
provided large part of the energy is contained in sufficiently  high frequencies.
The amplitude is not required to be small (apart from the shallow water approximation assumption)
and the near-linear behavior occurs on a much longer time scale than might be anticipated based
on the amplitude size.
Heuristically speaking, this effect is due to the nonlinearity getting averaged by the dispersive action.

This result is obtained by an averaging procedure, which is briefly outlined, and
is also confirmed by numerical simulations.
\end{abstract}

\pacs{ PACS 47.10.Fg, 47.35.Bb }
\maketitle


The dynamics of surface water waves has been an important object of study in science for over a century. Soliton solutions
and integrability in P.D.E.'s are two examples of remarkable discoveries that were made by
investigating  water wave dynamics in shallow waters. In more recent times,  the so-called rogue waves
have been under an intense investigation, see  for example \cite{KharifPelin2,osborne,zakharov} and the
references therein.
These unusually large waves
have been observed in various parts of the ocean in both deep, see {\em e.g.} \cite{KharifPelin}, and shallow water, see {\em e.g.} \cite{sand},
 leading scientists to suggest various mechanisms for rogue wave formation.

In the case of shallow water, one normally does not work with the full water wave equation but uses
approximate models to study the evolution, in particular rogue waves.
These models are nonlinear dispersive
equations such as KdV, Boussinesq approximations, {\em etc}. In particular, KdV describes
unidirectional small
amplitude long waves on fluid surface. See, {\em e.g.} \cite{talipova} for applications of KdV to
rouge waves in shallow water. Since rouge waves correspond to concentration of energy on small
domains, it is expected that high frequencies play important role in rouge waves formation.

In this Letter, we demonstrate that for sufficiently  high frequency initial data, one-dimensional
spatially periodic surface waves in  shallow water exhibit near-linear behavior. This phenomenon is
established by observing that high frequency periodic solutions in KdV equation  behave linearly.

Thus, linear theories of rogue wave formations can be extended
to nonlinear high frequency regime. While, we demonstrate this phenomenon for KdV, we expect
that similar behavior will occur for other models describing shallow water waves.
We note that our near-linear dynamics has nothing to do with weakly nonlinear theories as
we do not require small amplitude or small energy. The evolution is near-linear on larger time-scale
than one might expect based on the size of nonlinearity.
Moreover,  we deal with dynamics on bounded domains (periodic boundary conditions) so the solutions cannot scatter
to infinity. Near-linear behavior for high frequency solutions on unbounded
domains is easier to establish as, essentially, the energy disperses to infinity and
weakly nonlinear theories become applicable.

By contrast, on bounded domains ({\em e.g.} periodic, Dirichlet, or Neumann boundary conditions), near-linear behavior may occur only if the net effect
of nonlinearity  somehow averages out, as the nonlinearity stays of finite strength.
Heuristically, one may speculate that strong dispersion will cause high harmonics to
oscillate rapidly and average the nonlinearity out, assuming that there is no focusing
collapse in the problem. While such explanation seems plausible, it is important to formulate precisely
under what conditions and on what time scales such near-linear dynamics takes place and
this is what we undertake in this work.

Clearly, one has to be careful when considering short wave solutions for the equations obtained in
the long wave approximations such as KdV. However, we show that there is a  set of
parameters when our high frequency solutions correspond to a realistic physical scenario
in shallow  water waves, see Section \ref{sec:numerics}.

\section{Near-linear dynamics and rogue waves}

Rogue waves are large-amplitude waves appearing on the sea surface seemingly ``from nowhere''.
Such abnormal waves have been also observed in shallow water and KdV has been used to explain this
phenomenon \cite{talipova}.
In the oceanographic literature, the following amplitude criterion for the rogue
 wave formation is generally used: the height of the rogue wave should
exceed the significant wave height by a factor of 2-2.2 \cite{KharifPelin}. (Significant wave height
 is the average wave height of the one-third largest waves.)

There is a vast literature on rogue (also called freak and giant)  waves, as discussed above,
and many explanations have been proposed.  Major scenarios and explanations involve
\begin{itemize}
\item probabilistic approach: rogue waves are considered as rare events
in the framework of Rayleigh statistics
\item linear mechanism: dispersion enhancement (spatio-temporal focusing)
\item nonlinear mechanisms: in approximate models ({\em e.g.} KdV), for some special
initial data, large amplitude waves can be created.
\end{itemize}

Linear mechanism is very attractive as there are simple solutions leading to large amplitudes,
while nonlinear mechanism requires rather special initial data,
{\em e.g.} leading to the soliton formation. On the other hand, linear equations arise in
the small amplitude limit which is too restrictive.

Using near-linear dynamics in KdV one can experiment with a new mechanism of large wave formation
that combines linear and nonlinear
deterministic mechanisms.  Our results indicate that for a special but relatively large set
of initial data  (characterized by  the energy contained mostly in high frequency Fourier modes),
the solutions of KdV equation behave near-linearly. It is then possible {\em to construct large
amplitude solutions using linear mechanisms of large wave formation}.

Here, we illustrate our approach with periodic boundary conditions. This is not the most
realistic choice but appropriate for a model problem to illustrate the concept.
Indeed, while the sea surface is not periodic, one observes similar patterns over large areas.
It should be possible to extend near-linear dynamics to quasiperiodic and random initial data and
we intend to do it in the future investigation.

It would be also desirable to observe near-linear dynamics for
the full water wave problem, however, it is a considerably harder problem which
we also hope to address in our future work.

The fact that nonlinear dispersive systems exhibit near-linear behavior seems to be
rather general and should be important in other fields where nonlinear dispersive equations appear
as various approximations, such as solid state physics, photonics, {\em etc.}.

\section{Mechanism of near-linear dynamics in nonlinear dispersive systems}
Nonlinear dispersive equations that are derived as short or long wave, weakly nonlinear
approximations of basic physical models, often take the form
\[
u_t = L u + f(u)
\]
for a field variable $u$ which may represent slowly varying amplitude, rescaled velocity, {\em etc}.
The operator $L$ is a linear differential operator and $f$ is nonlinearity which may also contain
derivatives but of lower order than those  in $L$.

On infinite domains, dispersion tends to spread localized waves whereas nonlinearity may cause
shocks or focusing.
Here, as discussed in the introduction, we consider the phenomenon of {\em dispersive averaging}
when the dynamics occurs in a finite region and
dispersion cannot cause energy to spread to infinity. It turns out that dispersion creates rapid
oscillations
for high harmonics which effectively average out nonlinearity. If most of the energy is contained in
high frequencies, then $Lu$ will dominate over $f(u)$ assuming that we consider
subcritical nonlinearity.
However, it is not clear what is the effect of  the nonlinearity
over large times since $u$ is not small. We show that the nonlinear effects are small leading to essentially
linear evolution for finite but large times.

Let
\[
u = e^{Lt} v,
\]
so that
\[
v_t = e^{-Lt} f(e^{Lt}v).
\]
If $e^{Lt}$ is periodic in time then one can attempt to apply the usual averaging procedure directly. However,
it is not clear at all that the averaging procedure applies in this setting. Indeed, the above system
is a partial differential equation and also there is no clear separation of time scales.
Nevertheless, our main result described in the next section and subsequent numerical simulations
indicate that it could be  possible
to get around these difficulties and show that the above system is
close to the averaged one
\[
v_t = \langle e^{-L\tau} f(e^{L\tau}v) \rangle_{\tau}.
\]
Now, we describe this mechanism for KdV but, as we have mentioned, it seems to be applicable
with appropriate modifications to many other systems.

\section{Near-linear dynamics in KdV}
Consider KdV
\[
u_t = 6 uu_x + u_{xxx}
\]
with periodic boundary conditions $u(x+2\pi) = u(x)$.
The following statement characterizes near-linear dynamics in this equation: \\ \\

\begin{em}
Assume that initial data $u(x)$ has zero mean $\langle u \rangle = 0$ and has energy equal to 1:
\[
\int_{-\pi}^{\pi}u^2(x) dx = 1
\]
and assume that Fourier transform $\hat u$ satisfies
\[
\sum_{n\neq 0} \Big |\frac{\hat u(n)}{\sqrt{n}} \Big|^2 \leq C \epsilon^2.
\]

Then,
\[
\left (\int_{-\pi}^{\pi} |u(x,t)-e^{Lt}u(x,0)|^2 dx \right )^{\frac{1}{2}} \leq C_2 (\epsilon^2 + t \epsilon),
\]
where $e^{Lt}$ is the free propagator for KdV.
\end{em}

\vspace{5mm}

The proof of this statement will appear elsewhere and here we only present heuristic arguments.

In Fourier domain,
\[
u(x) = \sum_n v(n) e^{inx}
\]
the equations take the form
\[
\dot v(n) = in^3 v(n) + 6 \sum_{n_1+n_2 = n} n_1 v(n_1) v(n_2).
\]
After a time dependent change of variable
\[
v(n) = e^{in^3 t} w(n)
\]
we obtain
\[
\dot w(n) = 6 \sum_{n_1+n_2=n} e^{in_1 n_2 n t} n_1 w(n_1) w(n_2).
\]
The goal is to show that $w$ stays almost constant for large times under our
 high frequency assumptions
on the initial data. It is easy to see that for general solutions, one cannot expect this kind of
behavior and
on the other hand, we cannot estimate rate  of change of $w$ by neglecting averaging effects of
the exponent.
Indeed, without  the exponential factor  $e^{in_1 n_2 n t}$, the above system corresponds to Burgers equation
$u_t = uu_x$ which is known to have shocks and is known to exhibit strongly nonlinear dynamics.

Using high frequency assumptions and averaging effect of the exponent, we find that near-linear
dynamics can be established by applying a version of the normal form procedure,
see {\em e.g.} \cite{Kapeller}.  The detailed analysis is rather subtle and will appear elsewhere. Instead, we will
give an outline of the analysis and we will present results of our numerical simulations.

There are three types of terms in the above equations:

1. Low harmonics with $n_1,n_2$ small. They give small contribution  because $w$ is small in a certain
norm and there are few such terms.

2. High frequency harmonics with $n_1,n_2$ large. They are well averaged by the exponent.

3. Intermediate terms. There are not so many intermediate terms as the Diophantine
equation $n_1n_2n = N$ has few solutions ( on the order of $\log N$).

As a result, when one carries out two normal form transformations, which correspond
to changing  variable $w$ so as to absorb the action of non-resonant terms, one obtains
the system that indeed undergoes small change during the evolution.

\section{Numerical simulations and physical interpretation}
\label{sec:numerics}

The KdV equation has been used to describe surface water waves in the small
amplitude limit of long waves in shallow water. More precisely, two  parameters
are assumed to be small and equal
\[
\frac{\rm amplitude}{\rm depth} \sim
\left ( \frac{\rm depth}{\rm wave length} \right )^2 \ll 1.
\]
Therefore, one must be careful when considering high frequency
limit in KdV as it may have little relation to the actual wave dynamics.
However, in our case the near-linear dynamics phenomenon occurs when small
parameter $\epsilon$  characterizing high frequency limit (see the formula below),
is only moderately small $\epsilon = 0.4$, as numerical simulation show (one observes near-linear
dynamics for finite time).

\begin{figure}[b]
\begin{tabular}{cc}
\includegraphics[width=45mm]{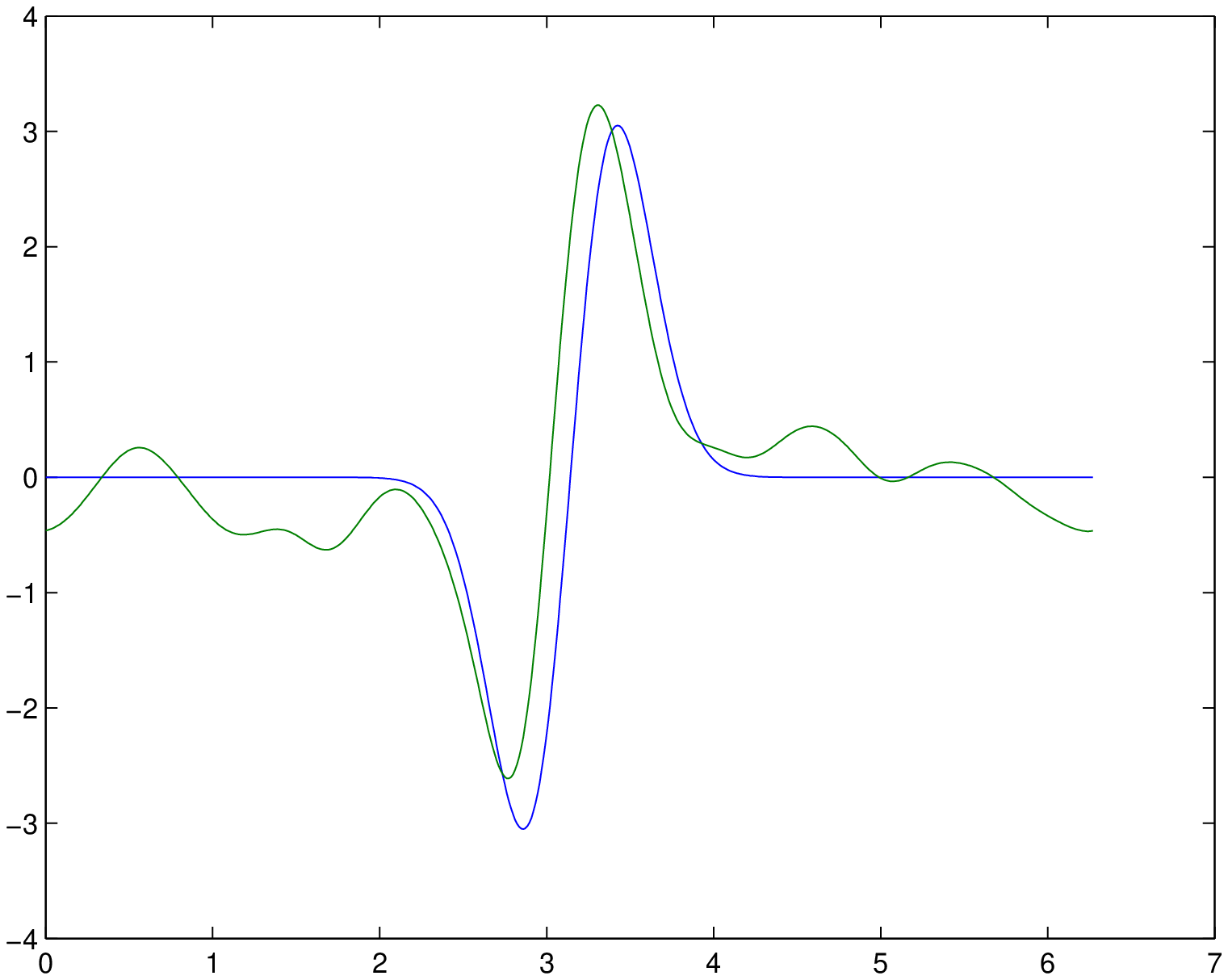} &
\includegraphics[width=45mm]{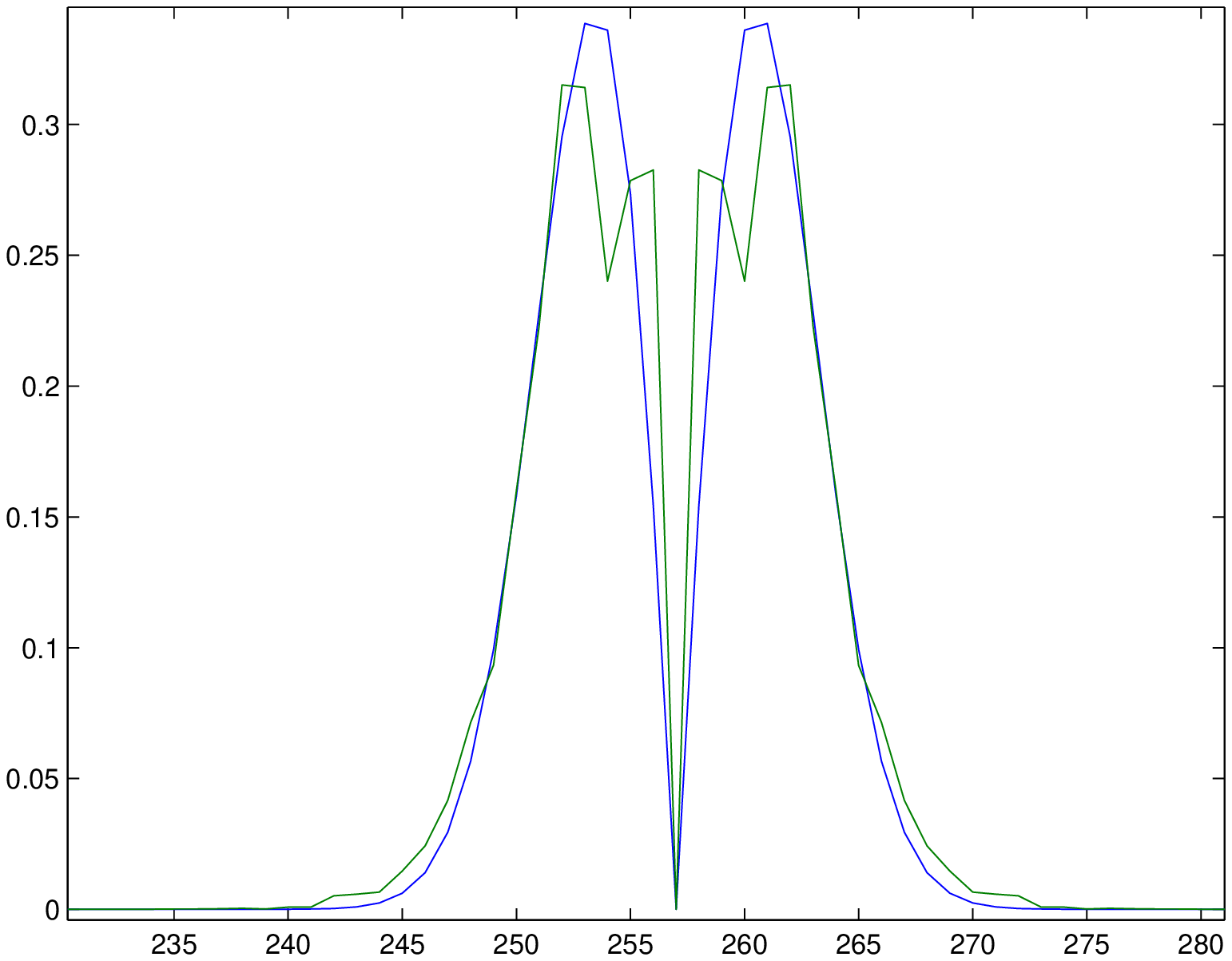}
\end{tabular}

\caption{Initial and evolved waves in KdV in Fourier space (left) and
 physical space (right).
Blue curve represents initial data and green curve represents
nonlinearly evolved data
with subsequent backward in time linear evolution, $e^{-Lt}u(x,t)$.
 Abscissa shows the number of
Fourier harmonic (left) and spatial coordinate (right),
while ordinate is the amplitude. The time of evolution is $T =1$.}
\label{fig1}
\end{figure}

\begin{figure}[b]
\includegraphics[width =60mm]{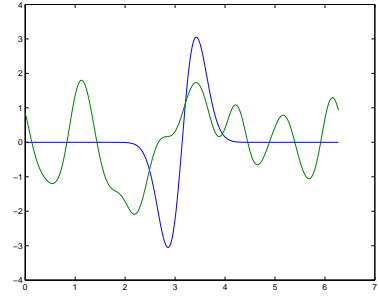}
\caption{This figure shows the initial data and the solution after $t=0.2$.
Because the initial data is spectrally broad, the initial wave
disperses over the whole periodic domain.}
\label{fig2}
\end{figure}

As initial data we take the scaled 1st Hermit function
\[
u(x) = \frac{4.5}{\sqrt{\epsilon}}\left ( \frac{x}{\epsilon} \right ) e^{-\frac{x^2}{2\epsilon^2}},
\]
so that the energy $\int u^2 dx$ does not depend on $\epsilon$
and is very close to 1.
For numerical simulations, we use KdV in the form
\[
u_t = \frac{3}{2} u u_x + \frac{1}{6}u_{xxx}
\]
as it appears in the derivation of KdV from the water wave equations (see below).
Specific numerical parameters are:  the length of periodic domain $L=2\pi$. The number of modes $M=2^9$.
Time step size $\Delta t = 10^{-7}$ with the time of the evolution $T = 1$.
 The discretization in space is given by $h =L/M$.
We used the so-called Fornberg-Whitham scheme which is described in
\cite{fornberg}.

Next, using standard derivation of KdV from water waves equations, we recall the
 relation between physical parameters and rescaled dimensionless variables, see
\cite{whitham}, Chapter 13.11.

Let $h_0$ be the depth when the water is at rest and let $Y = h_0+\eta$ be
the free surface of the water. Let $a$ be a characteristic amplitude and $l$ be a
characteristic wave length. Assume that
\[
\alpha= \frac{a}{h_0} \sim \beta = \frac{h_0^2}{l^2} \ll 1.
\]
Both $\alpha$ and $\beta$ are small parameters in the problem and they must
be of the same order.

Next, use the following natural normalization
\[
x^{\prime} = l x, \,\,\, Y^{\prime} = h_0 Y, \,\,\, t^{\prime} = lt/c_0, \,\,\, \eta^{\prime} = a \eta,
\]
where primed variables are the original ones and $c_0=\sqrt{g h_0}$.

The formal asymptotic expansion leads to KdV with higher order corrections
\[
\eta_t + \eta_x +\frac{3}{2}\alpha \eta \eta_x + \frac{1}{6} \beta \eta_{xxx}+
O(\alpha^2 + \beta^2)=0.
\]
Let  $X = x-t$ and $T = \alpha t$, so the equation becomes
\begin{equation}
\eta_T +\frac{3}{2}\eta \eta_X + \frac{1}{6}  \eta_{XXX}+
O(\alpha + \beta^2/\alpha)=0.
\label{eq:lasteq}
\end{equation}
One should expect that this approximation has accuracy of the order $O(\alpha)$ for finite
time $T = O(1)$, which implies $t \sim \alpha^{-1}$ and
$t^{\prime} \sim l/(c_0\alpha)$.

Finally, since we modify our solution with another parameter $\epsilon$, we verify that KdV
approximation will still make sense for some choice of the parameters.

First, let $\alpha=\beta = \delta \ll 1$.
Let us modify $a$ and $l$ with $ a_{\epsilon} = \frac{1}{\sqrt{\epsilon}} a$ and
 $l_{\epsilon} = \epsilon l$ which is consistent with our scaling of initial data. Then, we have

\[
\alpha_{\epsilon} = \frac{ a_{\epsilon}}{h_0} = \frac{\delta}{\sqrt{\epsilon}}, \,\,\,\,\,\,\,
\beta_{\epsilon} = \frac{h_0^2}{l_{\epsilon}^2} = \frac{\delta}{\epsilon^2}.
\]
These are small with $\delta = 0.01$ and $\epsilon = 0.4$. On the other hand the "mismatch" in the
equation (\ref{eq:lasteq}) is
\[
\alpha_{\epsilon}+ \frac{\beta_{\epsilon}^2}{\alpha_{\epsilon}} = \frac{\delta}{\sqrt{\epsilon}}+
\frac{\delta}{\epsilon^{3.5}} \approx \frac{1}{4}.
\]
Therefore, our high frequency regime may approximate water waves dynamics for example
with the following parameters: $a  =  1$ m, $h_0  =  100$ m, and
$l  =  1000$ m. \\

In summary,
we have observed a new mechanism of near-linear behavior in strongly nonlinear dispersive
equations. Precise formulations are given for KdV and implications for water waves
dynamics is found. \\

\begin{acknowledgments}
The authors were partially supported by NSF grants DMS-0600101 (B. E.), DMS-
0901222 (N. T.), and DMS-0807897 (V. Z.).
\end{acknowledgments}


\begin{thebibliography}{99}
\bibitem{KharifPelin2} C. Kharif, E. Pelinovsky, European Journal of Mechanics - B/Fluids,
Volume 22-6,  (2003) 603-634.
\bibitem{sand} S.E. Sand {\em et. al.}, Freak wave kinematics, in O. Torum, O.T. Gudmestad(Eds.), Water wave kinematics, Kluwer Academic Publishers, Dordrecht, (1990), pp. 535-549.
\bibitem{osborne} Alfred R. Osborne, Miguel Onorato, Marina Serio, Phys Lett A, 275, 5-6, (2000)
386-393.
\bibitem{zakharov} V.E. Zakharov, A.I. Dyachenko, A.O. Prokofiev, European Journal of Mechanics - B/Fluids, Volume 25-5, (2006) 677-692.

\bibitem{Kapeller} T. Kapeller, J. P\" oschel, KdV \& KAM, A Series of Modern
Surveys in Mathematics, Vol. 45, 2003.
\bibitem{newell} A.C. Newell, J.V. Moloney, Nonlinear Optics,  Advanced Topics in the Interdisciplinary Mathematical Sciences, Addison-Wesley, 1992.
\bibitem{KharifPelin} E. Pelinovsky, C. Kharif (Eds.), Extreme Ocean Waves,
Springer 2008.
\bibitem{talipova} E. Pelinovsky, T. Talipova, C. Kharif, Physica D 147 (2000) 83-94.
\bibitem{whitham} G.B. Whitham, Linear and Nonlinear Waves, Wiley, New York, 1974.
\bibitem{fornberg} B. Fornberg and G.B. Whitham, Phil. Trans. Roy. Soc. London A 289, 373 (1978).
\bibitem{sand} S E Sand, N E O Hansen, P Klinting, O T Gudmestad, M J Sterndorff,
PROC NATO ADVANCED RESEARCH WORKSHOP ON WATER WAVE KINEMATICS (1990)
Volume: 178, Pages: 535-549.
\end{thebibliography}
\end{document}